\documentclass[prl,twocolumn,showpacs,preprintnumbers,amsmath,amssymb]{revtex4}
\usepackage{graphicx}
\usepackage{dcolumn}
\usepackage{bm}
\begin{document}
\title{Towards ferromagnetic quantum criticality in FeGa$_{3-x}$Ge$_x$: $^{71}$Ga NQR as a zero field microscopic probe}
\author{M. Majumder$^{1}$}
\email{mayukh.cu@gmail.com}
\author{M. Wagner-Reetz$^1$}
\author{R. Cardoso-Gil$^1$}
\author{P. Gille$^2$}
\author{F. Steglich$^1$}
\author{Y. Grin$^1$}
\author{M. Baenitz$^1$}

\affiliation{$^1$Max Plank Institute for Chemical Physics of Solids, 01187 Dresden, Germany}
\affiliation{$^2$Ludwig-Maximilians-Universit$\ddot{a}$t M$\ddot{u}$nchen, Germany}

\date{\today}

\begin{abstract}

$^{71}$Ga NQR, magnetization and specific heat measurements have been performed on polycrystalline Ge doped FeGa$_3$ samples. A crossover from an insulator to a correlated local moment metal in the low doping regime and the evolution of itinerant ferromagnet upon further doping is found. For the nearly critical concentration at the threshold of ferromagnetic order, $x_C$=0.15, $^{71}(1/T_1T$) exhibits a pronounced T$^{-4/3}$ power law over two orders of magnitude in temperature which indicates 3D quantum critical ferromagnetic fluctuations. Furthermore, for the ordered x = 0.2 sample ($T_C\approx 6K$) $^{71}$($1/T_1T$) could be fitted well in the frame of Moriya's SCR theory for weakly FM systems with $1/T_1T \sim \chi$. In contrast to this, the low doping regime nicely displays local moment behavior where $1/T_1T \sim \chi^2$ is valid. For $T\rightarrow 0$ the Sommerfeld ratio $\gamma$=(C/T) is enhanced (70 mJ/mole-K$^2$ for for $x=0.1$) which indicates the formation of heavy 3$d$- electrons.

\end{abstract}
\pacs{76.60.-k, 75.50.Gg, 75.30.Et, 75.25.Dk}
\maketitle

Strongly correlated electron systems exhibit unconventional magnetic and electronic properties due to the presence of competing interactions. As a consequence of this competition combined with quantum fluctuations, the critical temperatures of phase transitions may continuously approach zero and a quantum critical point (QCP) emerges. QCPs are located therefore at zero temperature and they cause quantum fluctuations between competing ground states when a material is continuously tuned with a parameter (chemical pressure, mechanical pressure, magnetic field, etc.)\cite{Gegenwart08,Si10}. Recently spatial dimensionality of a material has been demonstrated as an insightful ingredient in tuning the degree of fluctuations and the physics of quantum criticality\cite{Custers12}. Quantum critical fluctuations cause non-Fermi-liquid (NFL) behavior and divergencies in physical properties, e.g. the effective charge carrier mass\cite{Stewart01}.

While to date, several antiferromagnetic (AF) QC systems have been studied extensively in 4$f$- and 3$d$- systems and the quest for ferromagnetic quantum criticality (FMQC) is of great interest\cite{Brando15}. This remains to be a less explored topic compared to AFMQCs, simply because FMQCPs are commonly avoided along two ways: It has been seen that pressure can decrease the FM transition temperature down to a reasonably low temperature, but just before reaching FMQCP the transition changes from second order to first order and the pure FMQCP is wiped out\cite{Pfleiderer97}. Isovalent or aliovalent substitutions can also systematically reduce the ordering temperature before some AF order or a more complex phase develops. Usually the substitution create strong disorder which forms then a Kondo cluster glass state\cite{Westercamp09} or other, more glassy states, like the Griffith phases for example\cite{Vojta10,Hu12}. Furthermore spiral spin arrangements\cite{Pedder13} or competing correlations (FM versus AFM)\cite{Sarrao98,Fernandez11,Ishida02,Gegenwart07} have also been found in some systems which prevents the formation of a pure FMQCP. It is therefore not surprising that a FMQCP has so far been verified only in one material i.e. the 4$f$ Kondo lattice system YbNi$_4$(P$_{0.9}$As$_{0.1}$)$_2$\cite{Steppke13}.

According to theoretical investigation in itinerant 3D and 2D systems, the FMQCP is unstable\cite{Kirkpatrick12,Chubukov04,Conduit09} and actually the system approaches to an incommensurate ordering or a first order into a commensurate state\cite{Kirkpatrick12,Chubukov04,Conduit09} as experimentally observed. According to Chubukov et. al. a system may undergo a Pomeranchuk instability into a p-wave spin-nematic state before a FM QCP is reached\cite{Chubukov04}.

Among 3$d$-intermetallic system the search for new quantum critical matter has various approaches. The first one was to study itinerant 3$d$ magnets at the verge of ferromagnetic order (NbFe$_2$\cite{Moroni09}, Ta(Fe,V)$_2$\cite{Horie10}), a second one was on diluted Fe-based systems(YFe$_2$Al$_{10}$\cite{Khuntia12}, YbFe$_2$Al$_{10}$\cite{Khuntia14}). A third approach was to study non magnetic 3$d$- (Fe, Co, Cr) based semimetals. Among them binary Fe based semimetals like FeSi, FeSb$_2$ and FeGa$_3$ earned great attention because of their strong correlations evolving at low temperatures accompanied by large thermopower peaks which makes them promising candidates for thermoelectrics\cite{Macro,Reetz14,Maik14}. Here metallic behavior and Fe based magnetism can be introduced by various substitutions. In contrast to the Te substitution in FeSb$_2$ (where Griffith phases evolve due to disorder)\cite{Hu12}, Ge doped FeGa$_3$ has been claimed to be a system with much less disorder\cite{Umeo12,Haldolaarachchige13} where pure FMQC might be observed. FeSb$_2$ and FeGa$_3$ itself are already at the verge of correlations as seen from NQR measurements where at low temperature fluctuation effects on $1/T_1$ have been seen and have been successfully described by the "correlated" in-gap states model for both systems\cite{Gippius142,Gippius14}. To explain the experimental phase diagram of electron or hole doped FeGa$_3$, theoretical calculations predicted that the systems turned into half-metallic upon electron or hole substitution\cite{Singh13}. Small Co substitution in FeGa$_3$ furnishes metallic behavior along with AFM correlations. Surprisingly these correlations resemble heavy fermion behavior which indicates a localization of Fe moments\cite{Gippius14}. Large Co-substitution however creates large disorder evidenced by broad Ga-NQR lines\cite{Gippius14}. Nonetheless it could be speculated that doped semiconductors opens up a new route for correlated $d$-electron physics. Fe(Ga,Ge)$_3$ seems to be an ideal platform to study the evolution of a metallic magnetic state and furthermore to study the system exactly at the putative FMQCP.

Here we employ nuclear quadrupolar resonance (NQR) on the two Ga- sites to explore the magnetic fluctuations in zero field on a microscopic scale over the entire phase diagram. There are mainly two important aspects about these studies. First NQR is zero field probe so the field effect on magnetic correlations (of RKKY type, Kondo type) is absent. The second important aspect is related to the local character of the method. Usually in NQR the line position is given by the local electric field gradient (EFG) and the line width is governed by local disorder and local transferred fields (under the presence of magnetic ion). Therefore information about local disorder and magnetic order is obtained simultaneously to the spin-fluctuations.

Two $^{71}$Ga NQR lines are found for each sample which corresponds to the two crystallographic Ga sites being present in the structure (Fig. 1(a)). The NQR line positions of the parent compound are in good agreement to the previous obtained result on powder material\cite{Gippius14}. The linewidth in Fig.1(a) is much smaller than the reported results\cite{Gippius14} due to the use of high quality single crystals. The nearly isotropic and narrow line width indicates the absence of strong local disorder in this series which is rather important to study FM critical phenomenon. As the Ge concentration increases the line width increases slightly. There is no considerable temperature dependence of the NQR frequency and linewidth for all samples. Figure 1(b) shows the temperature dependence for $x=0.1$ as an example. It should be mentioned that above the critical concentration $x_C\approx0.15$ the width gets broadened (see supplementary materials).

\begin{figure}[h]
{\centering {\includegraphics[width=0.5\textwidth]{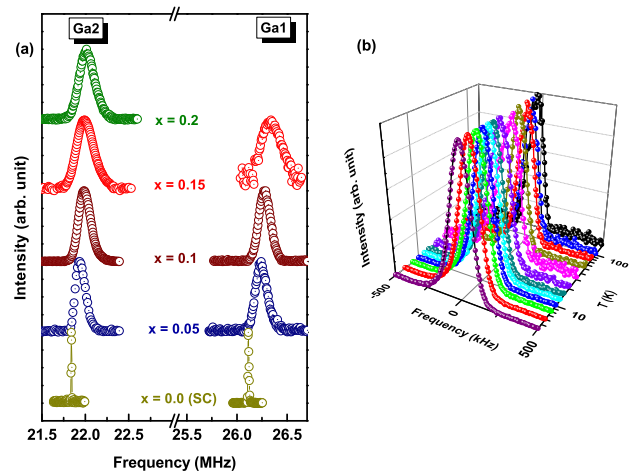}}\par} \caption{(Color online) (a): $^{71}$Ga NQR spectra for x=0.0 (single crystal), 0.05, 0.1 and 0,2 samples at 6K. (b): Temperature dependence of NQR spectra of Ga2 site for the $x=0.1$ sample around the center frequency of $\nu_0$= 22 MHz.} \label{structure}
\end{figure}


The nuclear spin-lattice relaxation time ($T_1$) has been measured for the $^{71}Ga$ isotope for both the Ga- sites to probe the evolution of critical fluctuations at zero field throughout the phase diagram. The temperature dependence of $1/T_1T$ have been plotted in figure 2(a) for the Ga2-site. Measurements are performed on both Ga- sites and $1/T_1T$ for the same sample shows a similar temperature dependence which indicates that both the Ga- site are experiencing same hyperfine field fluctuations (linearity of $1/T_1T$ (Ga1) versus $1/T_1T$ (Ga2) sites in Fig. S4\cite{Supplement}) but the magnitude is dissimilar which is due to the difference in hybridization of the Ga- orbitals with Fe 3$d$ orbitals (see supplement). The Ga2-site has three Fe neighbors whereas the Ga1-site has only two neighbors which might leads to a smaller $T_1$ value for the Ga2-site with respect to the Ga1-site. It seems from the crystal structure that the direction of the principle component of the electric field gradients ($V_{zz}$) are different at different Ga-sites. As $1/T_1$ in NQR probes the spin-fluctuations perpendicular to the direction of the principle component of the electric field gradients ($V_{zz}$), tell us that in our case electron spin-fluctuations are rather isotropic.

To model the temperature dependence of $1/T_1T$ we applied a two relaxation channel model with
\begin {equation}
1/T_1T = R_{3d} + R_{CE} = (1/T_1T)_{3d} + (1/T_1T)_{CE}
\end {equation}
where the first term derives from transferred Fe 3$d$ magnetic spin-fluctuations and the second term is the uncorrelated conduction electron contribution. At high temperatures the conduction electron term dominates which gives a $(1/T_1T)_{CE}$ = \textit{constant} term known as Korringa term. Here it could be speculated that upon doping with Ge at the Ga site, the density of states (DOS) at the Fermi level increases which lead to an increase in $(1/T_1T)_{CE}$ because $(1/T_1T)_{CE} \sim N(E_F)^2 \sim (C/T)_{CE}^2$. This is indeed the case for our samples(see table 2 in supplementary information). Towards low temperatures the term $(1/T_1T)_{3d} \sim \sum_q A_q^2 \chi^{\prime \prime}(q,\omega)$ dominates. The isotropic approach to model the low temperature upturn of $1/T_1T$ is a simple power law $(1/T_1T)_{3d} \sim C/(T^n-\Theta) \sim AT^{-n}$(if $\Theta$ goes to zero), frequently used for other correlated systems at the proximity to the magnetic order.

\begin{figure}[h]
{\centering {\includegraphics[width=0.5\textwidth]{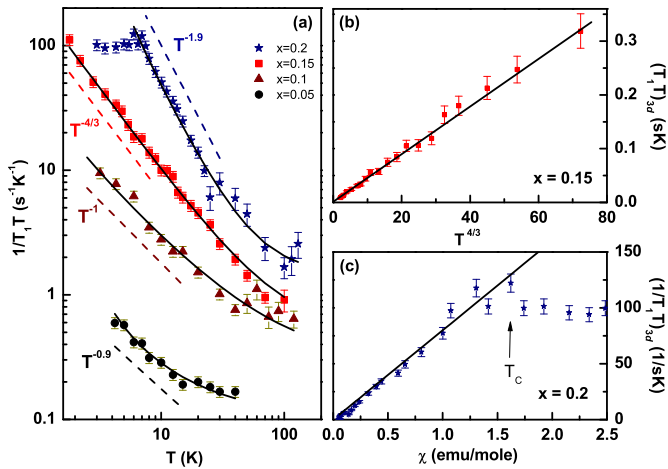}}\par} \caption{(Color online) (a): Temperature depndence of $1/T_1T$ for the Ga2 site and the dashed line fitted corresponds to the power law at low temperatures and the solid line corresponds to Eq. 1, (b) $(T_1T)_{3d}$ versus $T^{4/3}$ for $x=0.15$, (c): $(1/T_1T)_{3d}$ versus $\chi$ at $\mu_0$H = 0.005 T for $x=0.2$.} \label{structure}
\end{figure}

For the critical sample with $x=0.15$ the exponent equals $n = 4/3$ over two decades of temperature which is highly consistent with the existence of three dimensional ferromagnetic quantum critical fluctuations (Figure 2(a)) expected from the SCR theory\cite{Ishigaki96}. Furthermore it can be easily seen from the $(T_1T)_{3d}$ versus $T^{4/3}$ plot in Figure 2(b) that the dynamical spin-susceptibility ($\chi^{\prime \prime}(q,\omega)$) diverges almost at $T=0$ which gives evidence that the critical fluctuations are quantum in nature for the $x=0.15$ system. The coefficient of electronic specific heat $C(T)/T$, shows a logarithmic divergence(see Fig. S2(b)\cite{Supplement}), $ln(T_0/T)$ (with $T_0 \simeq 49 K$), which is expected near an itinerant 3D FMQCP. Below 1 K $C(T)/T$ seems to saturate which might indicate that the $x=0.15$ sample do not exactly match the FMQCP but in the higher doping side of exact FMQCP. Furthermore, the field dependence of C/T follows a scaling relation in the $C/T(0)-C/T(H)$ vs $T/B^{0.6}$ plot (see supplement) within the temperature range 1K to 5K and field range 0.01 T to 14 T which indicates quantum critical nature of spin fluctuations for $x=0.15$.

The temperature dependence of $1/T_1T$ in the paramagnetic region ($T > T_C$) of the ordered sample (x=0.2) is linear to the bulk susceptibility $\chi(T)$ (Fig. 2(c)) and also linear to $C(T)/T$\cite{Supplement}, both showing a $T^{-1.9}$ power law, which is a fingerprint of a weakly 3D itinerant ferromagnet predicted by SCR theory\cite{Moriya85,Majumder09,Majumder10}. The absence of a clear peak in $1/T_1T$ as well as in (C/T)\cite{Supplement} at the ordering temperature for x=0.2 indicates that the ordering might be not long range order or in such higher doping range there might be disorder. Below 6 K $C(T)/T$ deviates from power law behavior and tends to saturate\cite{Supplement}. It has also been seen that below the ordering temperature there is a onset of irreversible magnetic behavior seen in FC-ZFC magnetization measurements\cite{Supplement} for $x=0.2$ indicative of a disordered state and also the ZFC magnetization shows a peak at the ordering temperature which indicates the presence of weak AFM correlations on the top of dominant FM correlations (see supplement). Unusual states near ferromagnetic quantum criticality have been proposed by theory due to the competing interactions\cite{Nussinov09} and in our case may be the competition between dominating FM and weak AFM correlations induces such states near a FMQCP. Though the effect of disorder or phase separation in this high doping side can not be fully discarded (see supplementary materials).

For small concentrations $1/T_1T$ increase with $T^{-1}$ at low temperatures which indicates the formation of heavy electron out of the low density of carriers. This is confirmed by the increase of the Sommerfeld coefficient plotted as a function of temperature in Fig. 3(c). Similar to other heavy fermion systems $1/T_1T$ could be modeled \cite{Nakamura96,Buttgen86,Kuramoto00,MacLaughlin89,Cox85} for $T>T_K$,
\begin {equation}
1/T_1T \propto \chi(T)/\Gamma,
\end {equation}
with $\Gamma \propto T^{0.5}$ and where $\Gamma$ is the dynamic relaxation rate of the local moment (Fig. 3(a)). Furthermore $1/T_1T \propto (\chi)^2$ was found (inset of fig. 3(a)) which indicates that the Korringa law is valid and $\chi \sim 1/\sqrt{T}$ which is indeed the case (Fig. 3(b)). Furthermore the (C/T) value at the lowest temperature is about 70 mJ/mole-K$^2$ which signals heavy fermion behavior. The divergence of $\chi(T)$ and $C(T)/T$ at low fields can be suppressed by applying external magnetic field (Fig. 3(b) and (c))\cite{Supplement}.

Additional evidence for the change of correlations with Ge substitution has been provided by the Wilson ratio $R_w = \frac{\pi^2 R}{3C}\frac{\chi(0)}{\gamma}$, where R is the ideal gas constant, C is the Curie constant, $\chi(0)$ is the low temperature susceptibility, and $\gamma$ is the electronic specific heat coefficient. The $R_W$- values at 2 K for x=0.05, 0.1, 0.15, 0.2 are 3.9, 1.8, 130, 215 respectively. For the low doping systems (x = 0.05 and 0.1) a value of  $R_W$ around 2 is compatible with heavy fermion system and for the nearly critical (x=0.15) and the ordered system (x=0.2) the highly enhanced $R_W$ value indicates strong ferromagnetic correlations.

\begin{figure}[h]
{\centering {\includegraphics[width=0.5\textwidth]{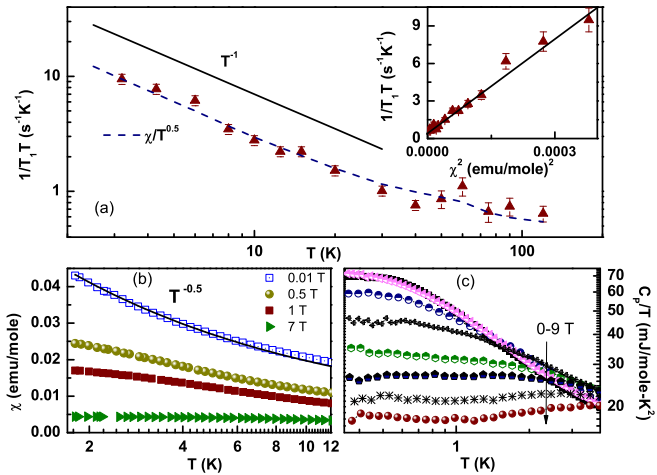}}\par} \caption{(Color online) (a): Temperature dependence of $1/T_1T$ for x=0.1, solid line indicate the power law of $T^{-1}$ and the dashed line is the fit according to equation 2 and the inset shows $1/T_1T$ versus (M/H)$^2$ at $\mu_0$H = 0.01 T with a solid line as a linear fit, (b) and (c): T dependence of $\chi=M/H$ and $C(T)/T$ at different fields respectively.} \label{structure}
\end{figure}

As a summery the full phase diagram of Ge doped FeGa$_3$ systems is plotted in Fig. 4. The low substituted Ge ($x\leq0.15$) systems exhibit a heavy fermion behavior. Increasing Ge concentration ($x\geq0.15$) promotes the evolution of ferromagnetic correlations and three dimensional quantum critical fluctuation have been confirmed for the $x=0.15$ sample. $1/T_1T$ at the lowest temperature (2 K) as a function of $x$ indicates a divergence of dynamical spin-susceptibility at the critical concentration, commonly found in those systems in the vicinity to the QCP. One can also estimate the critical temperature ($\Theta$(K)) where the dynamical spin susceptibility diverges (see Fig. 2(c), see supplementary fig. S8\cite{Supplement}), the $\Theta$(K) is $\simeq$ 0 for $x=0.15$ but for the other systems it has a finite value which indicates that the $x=0.15$ sample resides close to the FMQCP (Fig. 4). Larger substitutions of Ge ($x\geq0.2$) converts the short range ordering to long range ferromagnetic ordering. Upon increasing Ge content, $1/T_1T$ evidences an enhancement of both DOS and the correlations between Fe moments which induce long-range FM order (evident from the increment of $n$ in the lower panel of figure 4).

\begin{figure}[h]
{\centering {\includegraphics[width=0.5\textwidth]{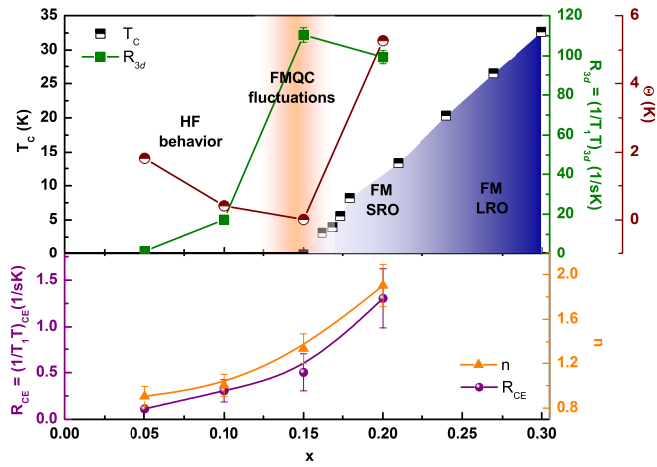}}\par} \caption{(Color online) Magnetic phase diagram ($T_C$- data taken from ref.\cite{Haldolaarachchige13}) and NQR parameters of our study as a function of Ge concentration. The half filled circles correspond to the $\Theta$ values.} \label{structure}
\end{figure}

In conclusion $^{71}$Ga NQR, magnetization and specific heat measurements have been performed in FeGa$_{3-x}$Ge$_x$ polycrystalline sample with $x = 0.05, 0.1$ (absent magnetic order), $x = 0.15$ (nearly critical) and $0.2$ ($T_C$$\approx$6 K). The nuclear quadrupolar resonance (NQR) spectra provide the evidence for the absence of strong intrinsic disorder due to Ge doping. Even more important the spin lattice relaxation rate provides the hyperfine field fluctuations on the Ga2- and Ga2- sites at zero field as an intrinsic measure of the Fe 3$d$ correlations. For the critical concentration of $x = 0.15$, $^{71}$($1/T_1T$) diverges at $T\rightarrow 0$ following a T$^{-4/3}$ power law over two orders of magnitude in temperature which indicates very pure 3D quantum critical FM fluctuations. For the $x = 0.2$ sample short range order is found and $^{71}$($1/T_1T$) could be fitted well in the frame of Moriya's SCR theory for weakly FM systems. In contrast to that low Ge substitutions in FeGa$_3$ cause the formation of heavy fermions as evidenced by the divergence of $C(T)/T$ towards lowest temperatures. The coexistence of correlated 3$d$ electrons and the ferromagnetic short range magnetic order in a very narrow stoichiometric range has not been seen before in other 3$d$- systems. We conclude that FeGa$_{3-x}$Ge$_x$ is a platform to study the rare occurrence of 3$d$ heavy fermions in the vicinity of a quantum critical point with very pronounced 3DFM fluctuations which is a rather unique scenario.

Recently, we became aware of a paper on Moessbauer measurements on Fe(Ga,Ge)$_3$. This local zero field method supports strongly the results presented here\cite{arxiv15}. Especially the presence of antiferro- and ferromagnetic spin density and, most important, localized Fe-moments are evidenced from their data.

We Thanks Prof. H. Yasuoka for fruitful discussions on the NQR part and M. Brando for discussions on the FM critical point. Furthermore discussions with H. Rosner and D. Kasinathan are acknowledged.

\end{document}